\documentstyle[aps,pra,epsf]{revtex}
\begin{document}
\draft
\title{Superconducting properties of the attractive Hubbard model}
\author{M. H. Pedersen$^1$\protect\thanks{To whom correspondence should be
    addressed. Present address: D\'epartement de Physique Th\'eorique,
Universit\'e de Gen\`eve, 1211 Gen\`eve, Switzerland.  E-mail:\ pedersen@serifos.unige.ch }
J. J. Rodr\'{\i}guez-N\'u\~nez$^2$, 
H. Beck$^3$,
T. Schneider$^1$\thanks{Present address: Physik-Institut der
Universit\"at
Z\"urich, 8057 Z\"urich, Switzerland}
and
S. Schafroth$^4$}
\address{$^1$IBM Research Division, Zurich Research
Laboratory, 8803 R\"uschlikon,
Switzerland 
\\
$^2$Instituto de Fisica, Universidade Federal Fluminense,
Av.\ Gal. Milton Tavares de Souza, 
Gragoat\'a S/N, \\ 24210-340 Niter\'oi RJ, Brazil. E-mail:\ jjrn@if.uff.br
\\
$^3$Institut de Physique, Universit\'e de Neuch\^{a}tel,
        2000 Neuch\^{a}tel, Switzerland
\\
$^4$Physik-Institut der Universit\"at Z\"urich,
8057 Z\"urich, Switzerland
}

\maketitle
\begin{abstract}
  A self-consistent set of equations for the one-electron self-energy
  in the ladder approximation is derived for the attractive Hubbard
  model in the superconducting state. The equations provide an
  extension of a $T$-matrix formalism recently used to study the effect
  of electron correlations on normal-state properties. An
  approximation to the set of equations is solved numerically in the
  intermediate coupling regime, and the one-particle spectral
  functions are found to have four peaks.  This feature is traced back
  to a peak in the self-energy, which is related to the formation of
  real-space bound states. For comparison we extend the moment
  approach to the superconducting state and discuss the crossover
  from the weak (BCS) to the intermediate coupling regime from the
  perspective of single-particle spectral densities.
\end{abstract}
\pacs{74.20.-Fg, 74.10.-z, 74.60.-w, 74.72.-h }

\clearpage
\section{Introduction}

High-temperature superconductors display a wide range of behavior
atypical of standard band-theory metals. In the superconducting state
these materials become extreme type II superconductors, with a very
short coherence volume, which one might take as an indication of
tightly bound pairs. Particularly interesting in this respect are
recent experiments showing a pseudo-gap structure in the 
normal-state density-of-states of underdoped
Bi$_2$Sr$_2$CaCu$_2$O$_{8+\delta}$ that 
persists almost up to room temperature \cite{ding,loeser}.
These features suggest that correlation effects
strongly affect the properties of such materials. One of the simplest
models featuring superconductivity and allowing a systematic study of
the effect of electron correlations is the attractive Hubbard model.
Although this model is unlikely to provide a microscopic description
of high-temperature superconductivity, it is likely that it reveals
the effect of correlations on measurable properties.

In a previous work, the effect of electron correlations on some
normal-state properties of the attractive Hubbard model was studied
using a self-consistent $T$-matrix formalism, going beyond simple
mean-field treatments \cite{tmatrix}. It was found that for
intermediate coupling strengths the attractive interaction gives rise
to large momentum real-space bound states with energies below the
two-particle continuum and with a pronounced effect on the spectral
properties, namely a splitting of the free band into two, one of them
associated with virtual bound states.  Also, a bending in the static
spin-susceptibility was observed for temperatures just above the
phase transition.

Thus, it is of significant interest to study the effect of 
these bound states also in the superconducting state. To this end we derive, in
Appendix A, a set of equations for the electron self-energies in the ladder
approximation including anomalous diagrams.  This is done in a general fashion
using the technique of functional derivatives, well-known from the textbook by
Kadanoff and Baym \cite{BK}. To facilitate a numerical analysis, the
anomalous part of the equations is approximated in a somewhat simpler form,
which is a reasonable approximation for the parameters of interest.  These
equations reduce to the familiar $T$-matrix equation in the normal state and
provide an obvious extension of the work embarked on in Ref.\ \cite{tmatrix}.
In Sec.\ \ref{eqsmot} we present a numerical solution using the
fast Fourier transform (FFT)
technique and discuss the physical meaning of the results.

To gain further insight into the electronic properties of the
superconducting state we compare the results with a generalization of
the moment approach, allowing a description of the diagonal and
off-diagonal single-particle spectral function in the superconducting
state, concentrating on the weak and intermediate coupling regimes.
This formalism is developed in Appendix B, and a comparison is
presented in Sec.\ \ref{momeqs}.

Our main results include (i) the appearance of four excitation
branches in the one-particle spectral function, with symmetric dispersions
around the chemical potential, and (ii) that the low-energy
behavior with respect to the chemical potential is in qualitative
agreement with BCS, whereas the high-energy features are not accounted
for in BCS.

\section{$T$-matrix equations below $T_{\rm c}$}\label{eqsmot}

The Hubbard Hamiltonian is defined as
\begin{eqnarray}
\label{Ham}
    H = - t\sum_{<ll'>\sigma}c_{l\sigma}^{\dagger}c_{l'\sigma}
   + U \sum_l n_{l\uparrow}n_{l\downarrow}
   - \mu \sum_{l\sigma} n_{l \sigma} \;\;\; ,
\end{eqnarray}
where $c_{l\sigma}^{\dagger}$($c_{l\sigma}$) are creation (annihilation)
electron operators with spin $\sigma$.  The number operator is $n_{l\sigma}
\equiv c_{l\sigma}^{\dagger}c_{l\sigma}$, $t$ a hopping matrix element
between nearest neighbors $l$ and $l'$, $U$ the on-site interaction,
and $\mu$ the chemical potential in the grand canonical ensemble.
Here we consider an attractive interaction, $U<0$. For a review
of the attractive Hubbard model, see Micnas et al.\ \cite{Micnas_et_al}.


In Appendix A we used functional derivatives to derive
self-consistent equations for the self-energy in the ladder approximation in the
superconducting state.  Here we present a numerical solution of a simplified
form of the derived equations, where the superconducting order parameter is only
taken into account to lowest order --- appropriate when it is significantly less
than the bandwidth.  The use of a ladder approximation means that only
two-particle scattering events are described and restricts the validity of the 
scheme to dilute systems. Thus, we approximated the self-energy by
\cite{Rodriguez,Beckapp}
\begin{equation}
    \Sigma(x,x') = \left( \begin{array}{cc}
              G_{22}(x,x')T(x',x) & \Delta(x)\delta(x-x') \\
            \Delta^*(x)\delta(x-x')  & G_{11}(x,x')T(x',x) \end{array} \right)
\;\;\; ,
\end{equation}
where $x=(l,\tau)$ denotes a space-time coordinate and 
the $T$-matrix in Fourier--Matsubara space is given by
\begin{equation}
    T(q,i\varepsilon_m) = \frac{U}{1-U \chi(q,i\varepsilon_m)}
\end{equation}
with
\begin{equation}
     \chi(q,i\varepsilon_m) \equiv \frac{1}{N\beta}
    \sum_{k,i\omega_n} G_{22}(k-q,i\omega_n-i\varepsilon_m)
     G_{11}(k,i\omega_n) \;\;\; ,
\end{equation}
where $i\varepsilon_m$ and $i\omega_n$ are bosonic and fermionic Matsubara
frequencies, respectively.
Another approximation approach, taking the order parameter into account in a
higher order than here, can be found in Ref.\ \cite{schafroth}.

In this formalism the two-particle Green's function,
which is also the local-pair propagator, is given by \cite{tmatrix}
\begin{equation}
    G_2({\bf k},i\varepsilon_n) = \frac{T({\bf k},i\varepsilon_n) 
        \chi({\bf k},i\varepsilon_n)}{U} \;\;\;.
\end{equation}

Green's functions are determined self-consistently using Dyson's
equation, Eq.\ (\ref{dyson}), and the order parameter is determined by
$\Delta(x)=U G_{12}(x^+,x)$. Technical aspects of the numerical
solutions using the FFT technique have been described in detail in Ref.\ 
\cite{tmatrix}.

The numerical simulations were performed in two dimensions (2D) for
$U/t=-6$, a density of $\rho=0.1$ and a temperature of $T/t=0.05$,
which is far below the transition temperature.  Here one must note
that, owing to the Mermin--Wagner theorem, no phase transition is expected
in 2D in systems with a continuous symmetry, and the formalism employed
is too simple to describe a Kosterlitz--Thouless phase transition.
Nonetheless, it is observed that the formalism does yield a
phase transition, in agreement with the results above the transition
temperature, where a divergence of the $T$-matrix was observed with
decreasing temperature.

\begin{figure}[h]
\vspace*{1cm}
\epsfysize=8cm
\centerline{\epsffile{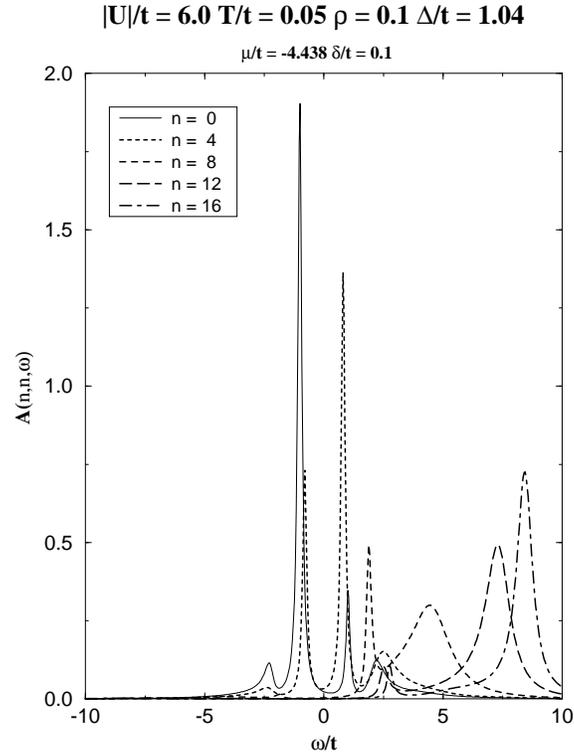}}
\vspace*{1cm}
\caption{ \label{fig1}
  Diagonal one-particle spectral function, $A({\bf k},\omega)$ vs.\
  $\omega$ for various momenta along the diagonal of the Brillouin
  zone, ${\bf k} = (n,n)\pi/16$, for $U/t = 6.0$, $T/t = 0.05$, and
  $\rho = 0.1$.  We have used an external damping of $\delta/t = 0.1$.
  After self-consistent calculation of the coupled nonlinear
  equations, we obtain the chemical potential, $\mu/t = -4.438$, and
  the order parameter, $\Delta/t = 1.04$.}
\end{figure}

In Fig.\ \ref{fig1}, we present the diagonal one-particle
spectral function, $A({\bf k},\omega)$, defined by
\begin{equation}
    A({\bf k},\omega) \equiv
    \lim_{\delta \rightarrow O^{+}}
     - \frac{1}{\pi} Im[G_{11}({\bf k},\omega + i\delta)],
\end{equation}
We see that four peaks exist for every momentum along the
diagonal of the Brillouin zone, a situation we will examine in a
moment.  The value of the order parameter is $\Delta/t=1.04$, which is
within the region of validity of the approximation, Eqs.\ 
(\ref{approx})--(\ref{self-energies}).  The transition temperature can
be found as the temperature where $\Delta$ vanishes and is $T_{\rm
  c}/t=0.2$, compared to the BCS transition temperature $T_{\rm
  c}^{\rm BCS}/t=0.66$. The zero-temperature order parameter, critical
temperature ratio is now $2\Delta(T=0)/k_{\rm B}T_{\rm c}=10.4$,
quite different from the canonical BCS result of $2\Delta(T=0)/k_{\rm
  B}T_{\rm c}=3.52$.

\begin{figure}[h]
\vspace*{1cm}
\epsfysize=8cm
\centerline{\epsffile{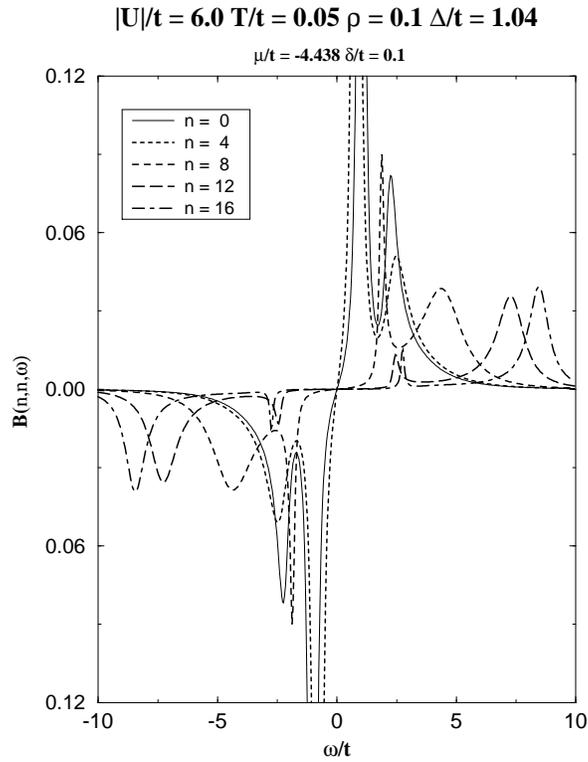}}
\vspace*{1cm}
\caption{ \label{fig2} 
  Off-diagonal one-particle spectral function, $B({\bf k},\omega)$
  vs $\omega$ for various momenta along the diagonal of the
  Brillouin zone.  Same parameters as in Fig.\  \protect\ref{fig1}.}
\end{figure}

In Fig.\ \ref{fig2}, we present the off-diagonal one-particle
spectral function, $B({\bf k},\omega)$, which is defined as
\begin{equation}
    B({\bf k},\omega) \equiv
    \lim_{\delta \rightarrow O^{+}}
    - \frac{1}{\pi} {\rm Im}[G_{12}({\bf k},\omega + i\delta)] \;\;\; .
\end{equation}
We also observe four peaks in the off-diagonal one-particle spectral
function which are at the same positions as the ones of the diagonal
one-particle spectral function. In Sec.\ III we will see that
symmetric poles around the chemical potential are a consequence of
Dyson's equation [Eq.\ (\ref{solu})].

\begin{figure}[h]
\vspace*{1cm}
\epsfysize=8cm
\centerline{\epsffile{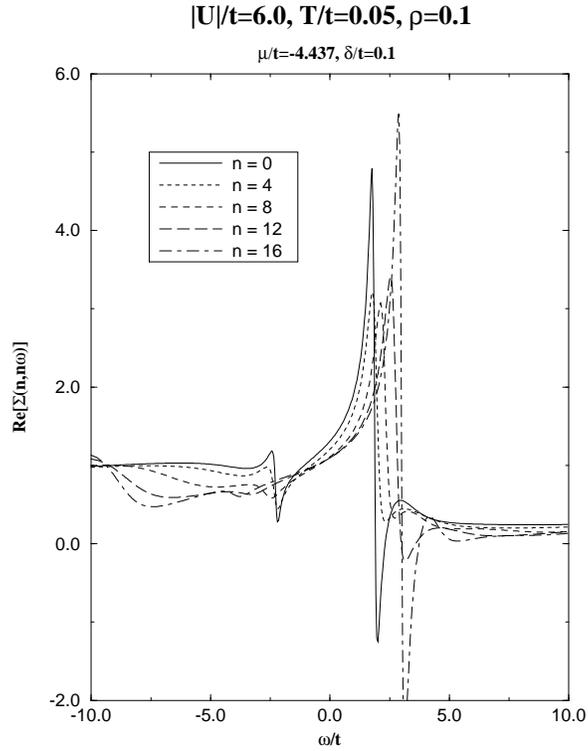}}
\vspace*{1cm}
\caption{ \label{fig3} 
  Real part of the self-energy, Re$[\Sigma({\bf k},\omega)]$ vs
  $\omega$ for various momenta along the diagonal of the Brillouin
  zone.  Same parameters as in Fig.\  \protect\ref{fig1}.}
\end{figure}

To explain the occurrence of these four peaks in the
one-particle spectral functions, let us study the real part of the
self-energy. This is presented in Fig.\ \ref{fig3}. Examining the
one-particle Green's function, $G({\bf k},i\omega_n)$, coming from
Dyson's equation, we find it is given by [see Eq.\ (\ref{solu})]
\begin{equation} 
\label{G11}
    G({\bf k},z)  = \frac{z + \varepsilon_{\bf k} +
    \Sigma({\bf k},-z)}{(z - \varepsilon_{\bf k} -
    \Sigma({\bf k},z))(z + \varepsilon_{\bf k} +
    \Sigma({\bf k},-z)) - |\Delta(T)|^2} \;\;\; 
\end{equation}
\noindent 
using $\Sigma_{22}({\bf k},z) = - \Sigma_{11}({\bf k},-z)$.
We wish to study the roots of the denominator of Eq.\ (\ref{G11}). For
example, in the BCS case where the self-energies are zero, we obtain
two solutions at $\omega = \pm \sqrt{\varepsilon_{\bf k}^2 +
  |\Delta(T)|^2}$.  For ${\bf k} = (\pi,\pi)$, we obtain
$\varepsilon_{\bf k}/t \approx 8.4$ because $\varepsilon_{\bf k}$ is
measured with respect to the chemical potential, which in our case is
$-4.438$ (see Fig.\ \ref{fig1}).  Then, the two roots are located at
$\approx \pm 8.5$.  These two roots are going to remain for the
interacting case, as we will see shortly. The positions of the poles
are determined mainly by the real part of the self-energy, and the
imaginary part contributes essentially only an additional lifetime effect.
Thus, we assume for simplicity that the imaginary part is zero. We then
have to find the roots of the equation
\begin{equation} \label{rootsU}
    (\omega - \varepsilon_{\bf k} - \Sigma'({\bf
    k},\omega))(\omega + \varepsilon_{\bf k} + \Sigma'({\bf k},-\omega)) -
    |\Delta(T)|^2 = 0 \;\;\; ,
\end{equation} 
where $\Sigma'({\bf k},\omega)$ is the real part of the
self-energy.  A rough estimate of the roots at ${\bf k} = (\pi,\pi)$ of
Eq.\ (\ref{rootsU}) yields the values $\omega/t \approx
\pm$2.6, $\pm$8.5.  We see that the two previous (i.e.\ BCS) solutions
remain and that there are two other solutions symmetric around the chemical
potential. These excitations correspond to electrons bound into local pairs.

\begin{figure}[h]
\vspace*{1cm}
\epsfysize=8cm
\centerline{\epsffile{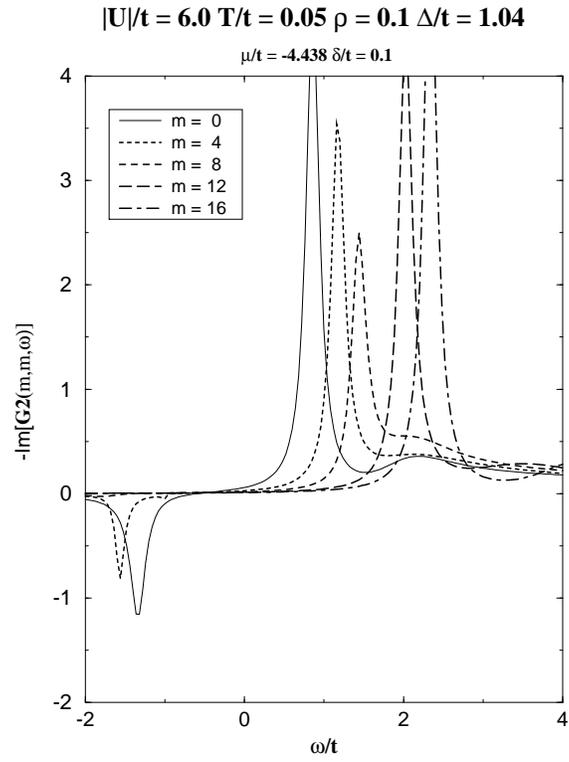}}
\vspace*{1cm}
\caption{ \label{fig4} 
  $-{\rm Im}[G_2({\bf q},\omega)]$ vs $\omega$ for different momenta along
  the diagonal of the Brillouin zone, ${\bf q} = (m,m)\pi/16$.  Same
  parameters as in Fig.\  \protect\ref{fig1}.}
\end{figure}

Figure \ref{fig4} shows the imaginary part of the two-particle Green's
function, $-{\rm Im}G_2({\bf q},\omega)$.  We observe that for zero momentum
we find the Cooper resonance, i.e., two symmetric peaks around the
chemical potential.  However, the Cooper resonance is different above
and below $T_{\rm c}$. Above the critical temperature the Cooper
resonance is sharp and close to the chemical potential, whereas in the
superconducting state the resonance is less sharp and the two peaks
are separated by an energy of $\approx 2\Delta$.  For larger momenta, we
find a peak that yields a narrow band (around $\omega \approx -2\mu +
U$).  This narrow band is associated with pair states, which is a
feature also observed in the $T$-matrix calculations above
$T_{\rm c}$ \cite{tmatrix}. The pair states give rise to a narrow band in the
self-energy, which in turn gives rise to a band of bound electrons in the
single-particle excitation spectrum. For this value of the
interaction, the band crosses that of the BCS quasiparticles,
causing a hybridization and correspondingly a split into four
excitation branches.

\section{Moment approach} \label{momeqs}

To further the understanding of the results, we have generalized the moment
approach\cite{Nolting} to the superconducting case. The formalism is
derived in Appendix B, and here we present numerical results and
compare them with 
the $T$-matrix results from above.

For the $T$-matrix formalism, we presented numerical data for
$U/t=6$ above. This interaction strength was chosen because it provides a
clear picture of the essential physics. Unfortunately, the moment
approach has a limitation
 for higher coupling strength, as will be discussed below, and we found it
necessary to restrict the coupling strength to $U/t=4$ in the
following.

\begin{figure}[h]
\epsfysize=10cm
\centerline{\epsffile{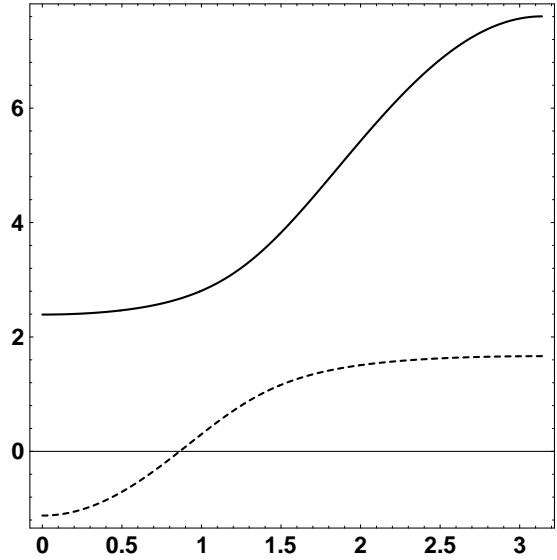}}
\caption{ \label{fig5}
Quasiparticle dispersion for the two poles in
the normal state at $T/t=0.2$ taken along the diagonal in the 
Brillouin zone for $U/t=4$, $\rho=0.1$.}
\end{figure}

\begin{figure}[h]
\epsfysize=10cm
\centerline{\epsffile{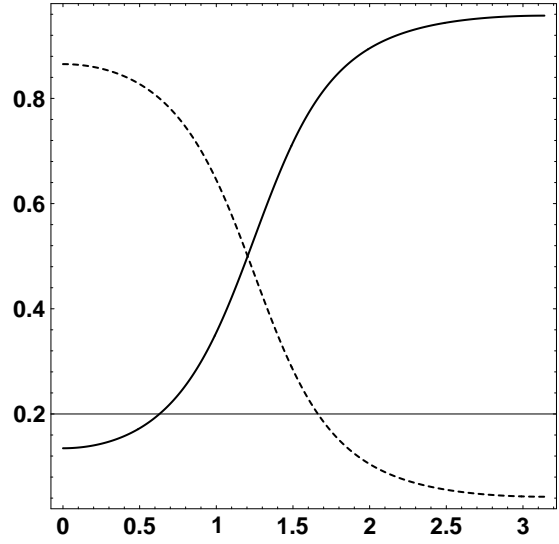}}
\caption{ \label{fig6} 
Amplitudes of the two poles in the normal state at $T/t=0.2$
taken along the diagonal in the Brillouin zone for $U/t=4$, $\rho=0.1$.}
\end{figure}

In Figs.\ \ref{fig5} and \ref{fig6} we show the dispersion of the
quasiparticles and their weights in the normal state for $U/t=4,
\rho=0.1$ in 2D.  It can be seen that the attractive interaction splits
the free quasiparticle band into two bands.  Examining the weights, we
identify the unpaired electrons as being the species of electrons with
dominating weights --- residing in the lower band for low momenta and in
the upper band for large momenta. With increasing $U$ the bands
separate more, and for some critical $U$ only paired electron reside in
the lower band, whereas only unbound electron reside in the upper
one \cite{moments}. Retaining this set of parameters for the moment, 
we calculated the dispersion of the quasiparticles and their weights in
the superconducting state using the moment approach outlined in
Appendix B.

\begin{figure}[h]
\epsfysize=10cm
\centerline{\epsffile{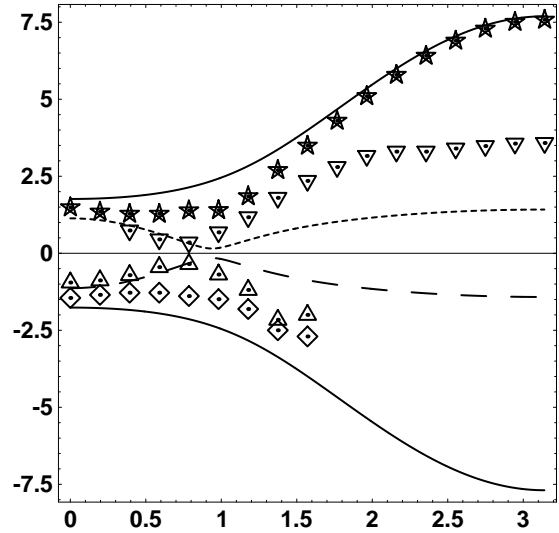}}
\caption{ \label{fig7} 
Quasiparticle dispersion for the four poles in the superconducting
state at $T/t=0.05$ taken along the diagonal in the Brillouin zone for $U/t=4$,
$\rho=0.1$.
For comparison, $T$-matrix results are included as symbols.}
\end{figure}

\begin{figure}[h]
\epsfysize=10cm
\centerline{\epsffile{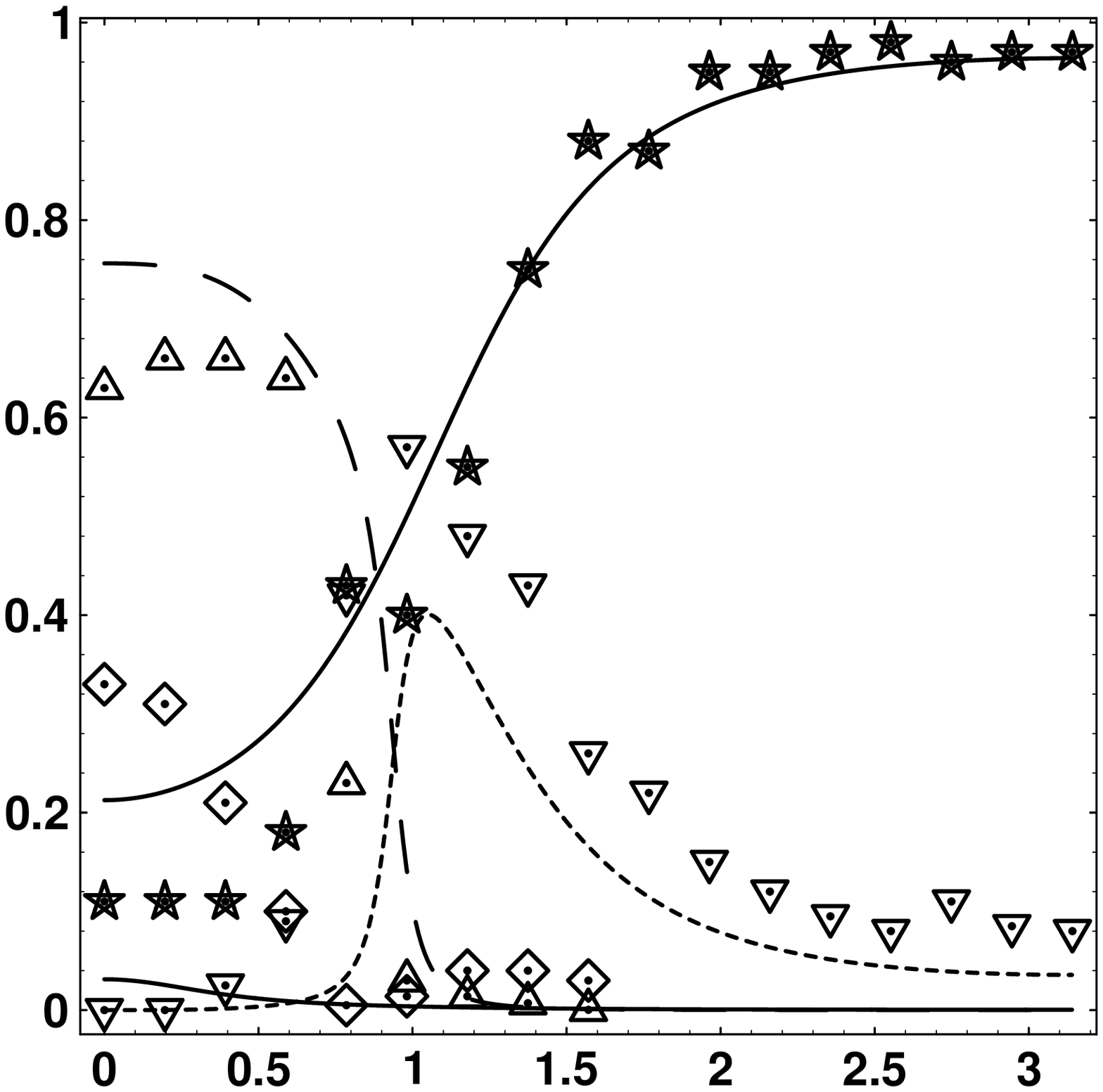}}
\caption{ \label{fig8} 
Amplitudes for the four poles in the superconducting state at $T/t=0.05$
taken along the diagonal in the Brillouin zone for $U/t=4$, $\rho=0.1$.
For comparison, $T$-matrix results are included as symbols.}
\end{figure}

The results for $U/t=4, \rho=0.1$ are shown in Figs.\ \ref{fig7} and
\ref{fig8}.  We obtain four bands instead of two in the
normal state, with a gap around the chemical potential. The lowest
band has negligible weight (Fig.\ \ref{fig8}) and is irrelevant for
the diagonal spectral function, although it makes a finite
contribution to the off-diagonal spectral function and thereby to the
gap equation and $T_{\rm c}$.  An essential point is that the band wherein
the chemical potential was located as well as the weight are
divided into two, leaving the upper band practically unchanged.  For
this reason, a BCS treatment still describes the behavior around the
chemical potential reasonably well, but misses the additional
excitation branches. These affect the off-diagonal Green's function
and hence the gap equation, Eq.\ (\ref{gabeq}). Indeed, the BCS
transition temperature $T_{\rm c}^{\rm BCS}/t=0.37$ 
is reduced to $T_{\rm c}/t=0.19$.
Thus, whereas the BCS transition temperature increased without bounds with
increasing coupling strength, the real transition temperature is reduced
compared to that resulting from the increasing pair mass. Indeed, for infinite
coupling strength all pairs would be localized and the transition temperature
would go to zero.
These features characterize the crossover from weak to intermediate
coupling. In Figs.\ \ref{fig7} and \ref{fig8} we have also included results
from numerical $T$-matrix calculations for the same coupling strength
and find quite a good overall fit. The discrepancy in the dispersion
for large momenta is also found in the normal state and can be
attributed to the neglect of the $k$-dependence in the approximation for
the band-correction term in Eq.\ (\ref{k-average}).

For increasing $U$ we find that the chemical potential moves into the
correlation gap and that the model becomes an insulator. 
The formalism used here yields
$T_{\rm c}=0$, which is correct in the sense that we no
longer expect a Fermi-surface instability for large $U$.
Instead we expect to have a scenario of Bose--Einstein condensation of
preformed pairs, yielding a finite transition temperature. The
description of this phenomenon is beyond the current formulation,
however. Considering that the moments describe the transition to the
strong-coupling regime well in the normal state, we expect that this
will also be so would we satisfy more off-diagonal moments. For instance
in our approach all odd off-diagonal moments are zero. This, however, is
not true for the exact moments. We thus believe that 
these nonzero moments must be satisfied in order to describe 
the crossover to Bose--Einstein condensation.

\section{Conclusions}\label{conclus}

Using a functional derivative technique, we have generalized the
particle--particle ladder approximation to the superconducting state,
providing an obvious extension of previous work in the normal state.
The resulting gap equation is a complicated, nonlinear integral
equation, whose solution is difficult to obtain.  Recently, 
Bickers and White evaluated the $T$-matrix
eigenvalues, which is equivalent to looking for normal-state
instabilities that signal second-order phase transitions \cite{BW}. In
contrast, the equations derived here cover the entire range of
temperature, both above and below $T_{\rm c}$.

Another calculation was carried out by Haussmann \cite{Hauss}, who
developed a perturbative formalism for the vertex function starting
from diagrammatic techniques, including both particle--particle and
particle--hole channels in the Bethe--Salpeter equation.  He ultimately
neglects the particle--hole channel, the vertex function is transformed
to the scattering $T$-matrix, and he treats superconductivity in a
mean-field fashion. Thus, no direct comparison with our equations is
possible.

A somewhat simplified version of the equations derived here has been
solved numerically using the FFT technique.  The main result is the
appearance of four peaks in both the diagonal and off-diagonal
one-particle spectral functions, which can be explained by a narrow
band in the self-energy, related to the formation of real-space bound
states below the two-particle continuum.

We presented an extension of the moment approach to the
superconducting state, providing an approximate expression of 
the single-particle spectral density. Although lifetime effects are
neglected and correlations effects in the band-correction term have
been treated approximately, our approach provides new insight into
the crossover from the weak to the intermediate coupling regimes.

For the specific case of $U/t=4,\rho=0.1$, we have shown that the
single-particle spectrum in the superconducting state exhibits four
symmetric branches with respect to the chemical potential. The low-energy
part (with respect to $\mu$) resembles BCS behavior,
whereas the high-energy physics is strongly modified.  This led to a
reduction of $T_{\rm c}$ in comparison with the expectation from BCS.

\section*{Acknowlegdments}
We thank the Swiss National Foundation for financial
support under project No.\ 21-31096-91 ({\it Theory of Layered
  Superconductors}).  One of the authors (JJRN) 
acknowledges partial support from the Brazilian Agency
CNPq (project no.\ 300705/95-6) and from CONICIT
(project no.\ F-139).  This work was carried out at IBM R\"uschlikon and
its kind hospitality is fully appreciated.  We thank R.\ 
Micnas and J.\ M.\ Singer for useful discussions.

\clearpage
\appendix

\section{Derivation of $T$-matrix equations}

It is convenient to define the Nambu--Green function
\begin{equation}
    {\cal G}(l\tau,l'\tau') =
    \left( \begin{array}{cc}
        -\langle T_\tau c_{l\uparrow}(\tau) 
c_{l'\uparrow}^\dagger(\tau') \rangle
        & \langle T_\tau c_{l\uparrow}(\tau) 
c_{l'\downarrow}(\tau') \rangle \\
        \langle T_\tau c_{l\downarrow}^\dagger(\tau) 
c_{l'\uparrow}^\dagger(\tau')
        \rangle &
        \langle T_\tau c_{l\downarrow}^\dagger(\tau) 
c_{l'\downarrow}(\tau') \rangle
    \end{array} \right)
\end{equation}
where $T_\tau$ is the time ordering operator.

The Dyson equation for the self-energies in Fourier--Matsubara space is

\begin{equation} \label{dyson} \label{GreenNam1}
   \left( \begin{array}{cc}
   i\omega_n - \varepsilon_{\bf k} - \Sigma_{11}({\bf k},i\omega_n) & -
    \Sigma_{12}({\bf k},i\omega_n) \\
    - \Sigma_{21}({\bf k},i\omega_n) & i\omega_n + \varepsilon_{\bf k} -
    \Sigma_{22}({\bf k},i\omega_n)
   \end{array} \right)
   \left( \begin{array}{ll}
   G_{11}({\bf k},i\omega_n) & G_{12}({\bf k},i\omega_n) \\
   G_{21}({\bf k},i\omega_n) & G_{22}({\bf k},i\omega_n)
   \end{array} \right)
        =
   \left( \begin{array}{rr}
   1 & {~0} \\
   0 & 1 \end{array} \right) \;\;\; ,
\end{equation}
\noindent 
where $i\omega_n \equiv \pi (2n+1)/\beta$ are the fermionic
Matsubara frequencies and $\beta=1/(k_{\rm B}T)$ the inverse temperature.
Formally the components of the Nambu--Green function can then be
expressed as

{\footnotesize
\begin{eqnarray} \label{solu}
    G_{11}({\bf k},i\omega_n)  &=& -G_{22}({\bf k},-i\omega_n) =
        \frac{i\omega_n + \varepsilon_{\bf k} -
        \Sigma_{22}({\bf k},i\omega_n)}{(i\omega_n - \varepsilon_{\bf k} -
        \Sigma_{11}({\bf k},i\omega_n))(i\omega_n + \varepsilon_{\bf k} -
        \Sigma_{22}({\bf k},i\omega_n)) - \Sigma_{12}({\bf k},i\omega_n)
        \Sigma_{21}({\bf k},i\omega_n)} ~ , \nonumber \\
    G_{12}({\bf k},i\omega_n) &=& G_{21}({\bf k},-i\omega_n) =
        \frac{\Sigma_{12}({\bf k},i\omega_n)}{(i\omega_n - 
\varepsilon_{\bf k} -
        \Sigma_{11}({\bf k},i\omega_n))(i\omega_n + \varepsilon_{\bf k} -
        \Sigma_{22}({\bf k},i\omega_n)) - \Sigma_{12}({\bf k},i\omega_n)
        \Sigma_{21}({\bf k},i\omega_n)} \;\;\; ,
\end{eqnarray}
}

\noindent
where the $d$-dimensional dispersion is given by
$\varepsilon_{\bf k} = -2t \sum_{\alpha} \cos(k_\alpha a_\alpha)-\mu$.
The problem of calculating the components of the one-particle
Nambu--Green function is then translated to the evaluation of the
self-energies \cite{Mattuck}.

The equation of motion for the Nambu--Green function contains
four-point correlation functions that cannot be evaluated
analytically. In order to derive a perturbation expansion we add auxiliary
source fields to the Hamiltonian
\begin{equation}
    H_{\rm S} = \sum_l \left(
    \rho_l^* c_{l\downarrow}^\dagger c_{l\uparrow}^\dagger
    +\rho_l c_{l\uparrow} c_{l\downarrow} \right)
\end{equation}
and ultimately take the limit of these fields going to zero.  This
particular way of generating four-point correlation functions leads in
the following to expressions for particle--particle scattering, which
in turn leads to superconductivity in the Cooper channel. It is
important to note that this approach does not generate particle--hole
diagrams.

The four-point correlation functions can now be written in terms of
derivatives with respect to the auxiliary fields, and the equation of
motion can be written 

\begin{eqnarray} \label{Dys(x,x')}
    \left( \begin{array}{ll}
    ~~~~~~~-\frac{\delta}{\delta \tau} + t_{ll'} + \mu & \Delta(x) 
\delta(x-x') +
    U \frac{\delta}{\delta \rho(x)} - \rho^*(x) \\
    \Delta^*(x) \delta(x-x') + U \frac{\delta}{\delta \rho^*(x)} + \rho(x) &
    ~~~~~~~~\frac{\delta}{\delta \tau}+t_{ll'} + \mu
  \end{array}\right)
 {\cal G}(x,x')
    = \left( \begin{array}{rr}
   1 & {~0} \\
   0 & 1 \end{array} \right) \delta(x-x') \;\;\; ,
\end{eqnarray}

\noindent 
where $\Delta(x) = U G_{12}(x^+,x)$ and $\Delta^*(x) = U G_{21}(x^+,x)$.
Here and in the following, we use the shorthand notation
$x=(l,\tau)$ for a space-time coordinate.
Then, from Eq.\  (\ref{Dys(x,x')}), the self-energy is given by
$\Sigma(x,x') = \Sigma^0(x,x') + \Sigma^1(x,x')$,
where
\begin{equation} \label{Dyson1's(x,x')}
    \Sigma^0(x,x') = \left( \begin{array}{cc}
                0 & \Delta(x) - \rho^*(x) \\
                \Delta^*(x) - \rho(x) & 0 \end{array} \right) \delta(x - x')
\end{equation}
and
\begin{equation} \label{Dyson2's(x,x')}
    \Sigma^1(x,\overline{a}) {\cal G}(\overline{a},x') = U \left( 
\begin{array}{cc}
                0 & \frac{\delta}{\delta \rho(x)} \\
                \frac{\delta}{\delta \rho^*(x)} & 0 
\end{array} \right) {\cal G}(x,x') \;\;\; ,
\end{equation}
where we used the notation
$f(\overline{a}) g(\overline{a})\equiv \int da f(a) g(a)$. In the
following, we also use the usual identity
$\delta {\cal G}(x,x') = {\cal G}(x,\overline{a}) \delta
\Sigma(\overline{a},\overline{b}) {\cal G}(\overline{b},x')$ \cite{BK}.

>From Eq.\ (\ref{Dyson2's(x,x')}) we find
\begin{equation} \label{sigm1vec}
    \Sigma^1(x,x') = U \sum_{s=1}^2 M_s {\cal G}(x,\overline{a})
    \frac{\delta \Sigma(\overline{a},x')}{\delta \rho_s(x)} \;\;\; ,
\end{equation}
where the matrices $M_{1(2)}$ are defined as 
\begin{equation} \label{Paulimat}
    M_1 =
            \left( \begin{array}{cc}
                0 & 1 \\
                0 & 0  \end{array} \right) \;\;\; ; \;\;\; 
M_2 =
             \left( \begin{array}{rr}
                0 & 0 \\
                1 & 0 \end{array} \right) 
\end{equation}
and $\rho_1(x)=\rho(x), \rho_2(x)=\rho^*(x)$.

We now introduce a {\it vertex function} 
\begin{equation}
    {\mit\Gamma}^{(s)}(x,x'|y) \equiv U \frac{\delta 
\Sigma(x,x')}{\delta \rho_s(y)} \;\;\; ,
\end{equation}
which obeys the equation

\begin{eqnarray} \label{vertex-tem}
    {\mit\Gamma}^{(s)}(x,x'|y) &=& - U\delta(x-x')\delta(x-y) M_s^{\rm T} 
\nonumber \\
    &+& U\delta(x-x') \left[ {\cal G}(x,\overline{a})
    {\mit\Gamma}^{(s)}
(\overline{a},
    \overline{b}|y) {\cal G}(\overline{b},x') \right]_{\rm off} \nonumber \\
    &+& \sum_{s'} M_{s'} {\cal G}(x,\overline{a}){\mit\Gamma}^{(s)}
(\overline{a},
    \overline{b}|y) {\cal G}(\overline{b},\overline{c}) 
{\mit\Gamma}^{(s')}(\overline{c},
    x'|x)  \nonumber \\
    &+& U\sum_{s'} {\cal G}(x,\overline{a}) \frac{\delta
    {\mit\Gamma}^{(s')}(\overline{a},x'|x)}{\delta \rho_s(y)} \;\;\; ,
\end{eqnarray}

\noindent
where $[...]_{\rm off}$ denotes the $2 \times 2$ matrix with
  zero diagonal elements and off-diagonal elements given by the matrix
  enclosed in brackets, and $M^{\rm T}$ is the transpose of the matrix
  $M$.  The self-energy is given by
\begin{equation} \label{self-good}
    \Sigma(x,x) = \Sigma^0(x,x') +
    \sum_s {\cal G}(x,\overline{a}) {\mit\Gamma}^{(s
      )}(\overline{a},x'|x) \;\;\; .
\end{equation}

As Eq.\ (\ref{vertex-tem}) is not practically manageable, we wish
to approximate it to a simpler form taking only ladder diagrams into
account, which describes repeated scattering between two particles.  This
contribution is expected to dominate for small densities.  A graphical
analysis will easily show that the latter term in Eq.\ 
(\ref{vertex-tem}) corresponds to higher-order diagrams, so
we omit it.  The eight matrix elements of
the vertex function then read

\begin{eqnarray} \label{vertexdyn}
    {\mit\Gamma}^{(s)}_{11} (x,x'|y) &=& G_{2 \alpha}(x,\overline{a})
        {\mit\Gamma}^{(s)}_{\alpha \beta} (\overline{a},
        \overline{b}|y) G_{\beta \gamma}(\overline{b},\overline{c})
        {\mit\Gamma}^{(1)}_{\gamma 1}(\overline{c},x'|x) \;\;\; , 
\nonumber \\
    {\mit\Gamma}^{(1)}_{12} (x,x'|y) &=& U \delta(x-x')G_{1 \alpha}
(x,\overline{a})
        {\mit\Gamma}^{(1)}_{\alpha \beta} (\overline{a},\overline{b}|y)
        G_{\beta 2}(\overline{b},x') 
     + G_{2 \alpha}(x,\overline{a}) {\mit\Gamma}^{(1)}_{\alpha \beta}
        (\overline{a}, \overline{b}|y) G_{\beta \gamma}
(\overline{b},\overline{c})
        {\mit\Gamma}^{(1)}_{\gamma 2}(\overline{c},x'|x) \;\;\; , 
\nonumber \\
    {\mit\Gamma}^{(1)}_{21} (x,x'|y) &=& - U \delta(x-x') 
\delta(x-y) + U \delta(x-x')G_{2
    \alpha}(x,\overline{a})
        {\mit\Gamma}^{(1)}_{\alpha \beta} (\overline{a},\overline{b}|y)
        G_{\beta 1}(\overline{b},x') \nonumber \\
    && +G_{1 \alpha}(x,\overline{a})
        {\mit\Gamma}^{(1)}_{\alpha \beta} (\overline{a},
        \overline{b}|y) G_{\beta \gamma}(\overline{b},\overline{c})
        {\mit\Gamma}^{(2)}_{\gamma 1}(\overline{c},x'|x) \;\;\; , 
\nonumber \\
    {\mit\Gamma}^{(s)}_{22} (x,x'|y) &=& G_{1 \alpha}(x,\overline{a})
        {\mit\Gamma}^{(s)}_{\alpha \beta} (\overline{a},
        \overline{b}|y) G_{\beta \gamma}(\overline{b},\overline{c})
        {\mit\Gamma}^{(2)}_{\gamma 2}(\overline{c},x'|x) \;\;\; , 
\nonumber \\
    {\mit\Gamma}^{(2)}_{12} (x,x'|y) &=& - U \delta(x-x') \delta(x-y)
    + 
U \delta(x-x')G_{1
        \alpha}(x,\overline{a}){\mit\Gamma}^{(2)}_{\alpha \beta}
       (\overline{a},\overline{b}|y) G_{\beta 2}(\overline{b},x') 
\nonumber \\
    && +G_{2 \alpha}(x,\overline{a}){\mit\Gamma}^{(2)}_{\alpha \beta} 
(\overline{a},
        \overline{b}|y) G_{\beta \gamma}(\overline{b},\overline{c})
        {\mit\Gamma}^{(1)}_{\gamma 2}(\overline{c},x'|x) \;\;\; , 
\nonumber \\
    {\mit\Gamma}^{(2)}_{21} (x,x'|y) &=& U \delta(x-x')G_{2 \alpha}
(x,\overline{a})
        {\mit\Gamma}^{(2)}_{\alpha \beta} (\overline{a},\overline{b}|y)
        G_{\beta 1}(\overline{b},x') 
    + G_{1 \alpha}(x,\overline{a}) {\mit\Gamma}^{(2)}_{\alpha \beta} 
(\overline{a},
        \overline{b}|y) G_{\beta \gamma}(\overline{b},\overline{c})
        {\mit\Gamma}^{(2)}_{\gamma 1}(\overline{c},x'|x) \;\;\; , \nonumber
\end{eqnarray}
\noindent
where summation on repeated indices is understood. Similarly, the
correction to the self-energy, $\Sigma^1$, is given by
\begin{eqnarray} \label{selfwvertex}
    \Sigma^1_{1\alpha}(x,x') &=& G_{21}(x,\overline{a}) 
{\mit\Gamma}^{(1)}_{1\alpha}
        (\overline{a},x'|x) + G_{22}(x,\overline{a}) 
{\mit\Gamma}^{(1)}_{2\alpha}
        (\overline{a},x'|x) \;\;\; , \nonumber \\
    \Sigma^1_{2\alpha}(x,x') &=& G_{11}(x,\overline{a}) 
{\mit\Gamma}^{(2)}_{1\alpha}
        (\overline{a},x'|x) + G_{12}(x,\overline{a}) 
{\mit\Gamma}^{(2)}_{2\alpha}
        (\overline{a},x'|x) \;\;\; . \nonumber
\end{eqnarray}

To further simplify this set of equations we expand it to second order in
the anomalous Green's functions, $G_{12}(G_{21})$, which yields the following
approximate form of the vertex functions
\begin{eqnarray} \label{approx}
    {\mit\Gamma}^{(s)}_{11} (x,x'|y) &\approx& G_{2 2}(x,\overline{a})
        {\mit\Gamma}^{(s)}_{21} (\overline{a},\overline{b}|y) G_{1
        2}(\overline{b},\overline{c}){\mit\Gamma}^{(1)}_{2 1}(\overline{c},
        x'|x) \;\;\; , \nonumber \\
    {\mit\Gamma}^{(1)}_{12} (x,x'|y) &\approx& U \delta(x-x')
        G_{1 2}(x,\overline{a})
        {\mit\Gamma}^{(1)}_{21} (\overline{a},
        \overline{b}|y) G_{1 2}(\overline{b},x')~~, \nonumber \\
    {\mit\Gamma}^{(1)}_{21} (x,x'|y) &\approx& - U \delta(x-x') \delta(x-y) + U
        \delta(x-x')G_{2 2}(x,\overline{a}) {\mit\Gamma}^{(1)}_{21}
        (\overline{a},\overline{b}|y) G_{1 1}(\overline{b},x') \;\;\;
        , \nonumber \\
    {\mit\Gamma}^{(s)}_{22} (x,x'|y) &\approx& G_{12}(x,\overline{a})
        {\mit\Gamma}^{(s)}_{21} (\overline{a}, \overline{b}|y)
        G_{11}(\overline{b},\overline{c}){\mit\Gamma}^{(2)}_{12}(\overline{c},
        x'|x) \;\;\; , \\  
    {\mit\Gamma}^{(2)}_{12} (x,x'|y) &\approx& - U \delta(x-x') \delta(x-y) + U
        \delta(x-x')G_{11}(x,\overline{a}){\mit\Gamma}^{(2)}_{12}
        (\overline{a},\overline{b}|y)G_{22}(\overline{b},x') \;\;\; , 
\nonumber \\
    {\mit\Gamma}^{(2)}_{21} (x,x'|y) &\approx& U \delta(x-x')
        G_{21}(x,\overline{a})
        {\mit\Gamma}^{(2)}_{12} (\overline{a},\overline{b}|y) G_{21}
(\overline{b},x')
        \;\;\; , \nonumber
\end{eqnarray}

>From Eqs.\ (\ref{approx}) we conclude that ${\mit\Gamma}^{(1)}_{21} 
(x,x'|y)$ and
${\mit\Gamma}^{(2)}_{12} (x,x'|y)$ are related to the $T$-matrix equations
above $T_{\rm c}$ \cite{tmatrix,Fresard}
\begin{equation} \label{vertex1-21} \label{2-12}
    {\mit\Gamma}^{(1)}_{21} (x,x'|y) = {\mit\Gamma}^{(2)}_{12} (x,x'|y) =
    -\delta(x-x') T(x,y) \;\;\; ,
\end{equation}
where $T(x,y)$ in reciprocal space has the form
\begin{equation} \label{T-matrix}
    T(q,i\varepsilon_m) = \frac{U}{1-U \chi(q,i\varepsilon_m)} 
\end{equation}
with
\begin{equation} \label{suscept}
     \chi(q,i\varepsilon_m) \equiv \frac{1}{N\beta}
    \sum_{k,i\omega_n} G_{22}(k-q,i\omega_n-i\varepsilon_m)
G_{11}(k,i\omega_n) \;\;\; ,
\end{equation}
\noindent 
and $i\varepsilon_n \equiv 2 \pi n/\beta$ are the bosonic
Matsubara frequencies.

Using Eqs.\ (\ref{vertex1-21}) and (\ref{T-matrix}), we obtain
the self-energies
to second order in $G_{12}(G_{21})$:

\begin{eqnarray} \label{self-energies}
    \Sigma^1_{11}(x,x') &=& G_{22}(x,x')T(x',x) +
        G_{21}(x,\overline{a}) G_{22}
(\overline{a},\overline{b})T(\overline{b},x)
            G_{12}(\overline{b},x')T(x',x)  \;\;\; , \nonumber \\
    \Sigma^1_{12}(x,x') &=& G_{22}(x,\overline{a})
G_{12}(\overline{a},\overline{b})
        T(\overline{b},x)G_{11}(\overline{b},x')
T(x',\overline{a}) \;\;\; , \nonumber \\
    \Sigma^1_{21}(x,x') &=& G_{11}(x,\overline{a})
G_{21}(\overline{a},\overline{b})
        T(\overline{b},x)G_{22}(\overline{b},x')
T(x',\overline{a}) \;\;\; , \nonumber \\
    \Sigma^1_{22}(x,x') &=& G_{11}(x,x')T(x',x) +
         G_{12}(x,\overline{a})G_{11}
        (\overline{a},\overline{b})T(\overline{b},x)
G_{21}(\overline{b},x')T(x',x)  \;\;\; .
\end{eqnarray}

This expansion is valid for $\Delta/W \ll 1$, where $W=2dt$ is the bandwidth.

To solve Eqs.\ (\ref{solu}) and
(\ref{T-matrix})--(\ref{self-energies}) one would also have to fix the
chemical potential from the particle number using
\begin{equation} \label{number}
    \rho(T,\mu) = \lim_{\eta \rightarrow 0^+}
 \frac{1}{\beta N}\sum_{\omega_n,{\bf k}}G({\bf k},i\omega_n)
       \exp(i\omega_n \eta) \;\;\; ,
\end{equation}
\noindent 
where $\rho$ is the electron concentration per spin and is
defined in the interval $[0,1]$.  Thus, the set of Eqs.\ (\ref{solu}) and
(\ref{T-matrix})--(\ref{number}) represents a set of nonlinear
self-consistent equations, which must be solved numerically.

We note that an expansion of the final equations,
Eqs.\ (\ref{approx})--(\ref{self-energies}), to first order in $U$ simply
yields the well-known BCS expressions. To second order in $U$, the result is
identical to that of Mart\'{\i}n--Rodero and Flores \cite{M-R&F}. The
second-order expansion was found to yield the same gap 
equation as in BCS, but with a renormalized interaction.

\clearpage
\section{Derivation of moment equations}

Introduce the diagonal and off-diagonal one-particle Green's function
\begin{equation}
        G_\sigma(i-j,\tau) =
         -\langle T_\tau c_{i\sigma}(\tau) c_{j\sigma}^\dagger(0)\rangle
\end{equation}
and
\begin{equation}
        F(i-j,\tau) = -\langle T_\tau c_{i\uparrow}(\tau)
        c_{j\downarrow}(0)
\rangle \;\;\; ,
\end{equation}
where $T_\tau$ is the time-ordering operator,
as well as the associated spectral functions
\begin{eqnarray}
        A(k,\omega) &=& -\frac{1}{\pi} {\rm Im} G(k,\omega+i\delta)
        \;\;\; ,
        \label{spectral_a} \\
        B(k,\omega) &=& -\frac{1}{\pi} {\rm Im} F(k,\omega+i\delta)
        \;\;\; .
        \label{spectral_b}
\end{eqnarray}
To construct approximate expressions for the spectral functions, it is
useful to consider the frequency moments
\begin{eqnarray}
    A_n(k) &=& \int_{-\infty}^\infty d\omega\, \omega^n A(k,\omega)
    \;\;\;, \\
    B_n(k) &=& \int_{-\infty}^\infty d\omega\, \omega^n B(k,\omega)
    \;\;\; .
\end{eqnarray}
With the $d$-dimensional dispersion defined as
\begin{equation}
        \epsilon_k = -2t \sum_{i=1}^d \cos(k_i r_i) \;\;\; , 
\end{equation}
the exact four first diagonal moments are given by\cite{Nolting}

\begin{eqnarray}
    A_0(k) &=& 1 \label{a0} \\
    A_1(k) &=& a_1 = \epsilon_k - \mu - \rho U \\
    A_2(k) &=& a_2 = (\epsilon_k-\mu)^2-2(\epsilon_k-\mu)\rho U+\rho U^2 \\
    A_3(k) &=& a_3 = (\epsilon_k-\mu)^3-3(\epsilon_k-\mu)\rho U+
        (\epsilon_k-\mu)U^2 \rho (2+\rho)
        + (\epsilon_k-\mu)\rho^2 U^2 \\
       && -U^3 \rho+U^2 \rho(1-\rho) K(k) \;\;\; , \label{a3}
\end{eqnarray}
where
\begin{eqnarray}
    \rho(1-\rho) K(k) &=& K+K_w(k) \label{kterm} \\
    K &=& \frac{1}{N} \sum_{ij} \langle c_{i\bar{\sigma}}^\dagger
        c_{j\bar{\sigma}}(2n_{i\sigma}-1)\rangle \label{k} \\
    K_w(k) &=& \frac{1}{N} \sum_{ij} t_{ij} e^{ik(r_i-r_j)} \{
        \langle n_{i\bar{\sigma}} n_{j\bar{\sigma}}\rangle-\rho^2 \\
        &-&
        \langle c_{j\sigma}^\dagger c_{j\bar{\sigma}}^\dagger
        c_{i\bar{\sigma}} c_{i\sigma}\rangle  - \langle c_{j\sigma}^\dagger
        c_{i\bar{\sigma}}^\dagger c_{j\bar{\sigma}}
        c_{i\sigma}\rangle \} \nonumber
\end{eqnarray}
and the first two off-diagonal moments are given by
\begin{eqnarray}
        B_0(k) &=& 0 \label{b0} \\
        B_1(k) &=& -U \sum_i \langle c_{i\downarrow}
        c_{i\uparrow}\rangle \;\;\; . \label{b1}
\end{eqnarray}

In the normal state the first four frequency moments have recently
been used in conjunction with an ansatz of the form
$A(k,\omega)=\tilde{\alpha}_1(k)\delta(\omega-\tilde{\Omega}_1(k))+
\tilde{\alpha}_2(k)\delta(\omega-\tilde{\Omega}_2(k))$.
The $k$-dependence in Eq.\ (\ref{kterm})
is approximated by its momentum average. Only Eq.\ (\ref{k}) contributes
to the average and is given by \cite{Nolting}
\begin{eqnarray} \label{k-average}
        \rho(1-\rho) K &=& -\frac{1}{N} \sum_k \sum_i
        \alpha_i(k) \epsilon_k n_f(\Omega_i(k)) \nonumber \\
        &\times&  \left[ \frac{2}{U}(\Omega_i(k)-\epsilon_k)+1 \right]
        \;\;\; ,
\end{eqnarray}
where $n_f(\omega)$ is the Fermi distribution function. The chemical
potential is obtained by fixing the density using
\begin{equation} \label{chempot}
        \rho = \sum_{k,i} \alpha_i n_f(\Omega_i(k)) \;\;\; .
\end{equation}

To analyze the superconducting state we make the assumption that the
general structure of the diagonal part of the self-energy does not change
in the superconducting state. As the diagonal part of the
self-energy is related primarily to pair-interaction physics, we expect
this to be a reasonable approximation. Thus, we express the
self-energy as \cite{tmatrix,moments}
\begin{equation}
        \frac{1}{z-\epsilon_k+\Sigma_\uparrow(k,z)} =
        \frac{\alpha_1(k)}{z-\Omega_1(k)} +
        \frac{\alpha_2(k)}{z-\Omega_2(k)} \;\;\; ,
\end{equation}
where $\alpha_1(k)$, $\alpha_2(k)$, $\Omega_1(k)$, $\Omega_2(k)$ are
functions to be determined. Note that this approximation neglects
lifetime effects.

The diagonal and off-diagonal Green's function in the superconducting state
is now given by Dyson's equation
\begin{eqnarray}
  G_\uparrow(k,z) &=& \frac{z+\epsilon_k-
        \Sigma_\downarrow(k,z)}
        {(z-\epsilon_k+\Sigma_\uparrow(k,z))
         (z+\epsilon_k-\Sigma_\downarrow(k,z))
         - |\Delta_k|^2} \\
  F(k,z) &=& \frac{\Delta_k}
        {(z-\epsilon_k+\Sigma_\uparrow(k,z))
         (z+\epsilon_k+\Sigma_\downarrow(k,z))
         - |\Delta_k|^2} \;\;\; ,
\end{eqnarray}

\noindent where $\Sigma_\downarrow(k,z)=-\Sigma_\uparrow(k,-z)$.
In Dyson's equation
 we have made the additional assumption that the off-diagonal part
of the self-energy, $\Delta_k$, is independent of frequency. This
assumption is likely to be wrong for large coupling strengths, where
$\Delta_k$ might contain additional structure related to the existence
of local pairs.

Combining the above, we can write Green's functions for frequencies
on the real axis as

\begin{eqnarray}
    G_\uparrow(k,\omega) &=&
        \frac{\alpha_1(k)(\omega+\Omega_1(k))(\omega^2-\Omega_2^2(k))+
        \alpha_2(k)(\omega+\Omega_2(k))(\omega^2-\Omega_1^2(k))}
        {(\omega^2-\omega_0^2(k))(\omega^2-\omega_1^2(k))}
        \label{green_g}  \\
    F(k,\omega) &=& \frac{
        \Delta_k [\omega^2-(\alpha_1(k)\Omega_2(k)+\alpha_2(k)\Omega_1(k))^2]}
        {(\omega^2-\omega_0^2(k))(\omega^2-\omega_1^2(k))} \;\;\; ,
\label{green_f}
\end{eqnarray}
where $\pm\omega_0(k)$ and $\pm\omega_1(k)$ are the dispersions
of the resulting four poles given by
{\footnotesize
\begin{equation}
        \omega_{0,1}^2(k)  =  \frac{1}{2} \left( \Delta_k^2+\Omega_1^2(k)+
        \Omega_2^2(k)  \pm
        \sqrt{[\Delta_k^2+\Omega_1^2(k)+\Omega_2^2(k)]^2
        -4[\Omega_1^2(k) \Omega_2^2(k) + \Delta^2_k
        (\alpha_1(k)\Omega_2(k)+\alpha_2(k)\Omega_1(k))^2]} \right)
      \;\;\; .
\end{equation}
}

The corresponding spectral functions,
Eqs.\ (\ref{spectral_a}) and (\ref{spectral_b}), can then be written as 
$A(k,\omega)=\tilde{\alpha}_1(k) \delta(\omega-\omega_0)+
\tilde{\alpha}_2(k) \delta(\omega+\omega_0)+
\tilde{\alpha}_3(k) \delta(\omega-\omega_1)+
\tilde{\alpha}_4(k) \delta(\omega+\omega_1)$ and
$B(k,\omega)=\beta_1(\delta(\omega-\omega_0)-\delta(\omega+\omega_0))+
\beta_2(\delta(\omega-\omega_1)-\delta(\omega+\omega_1))$.
The weights are found as the residues of the poles and given by
 
\begin{eqnarray}
    \tilde{\alpha}_1(k) &=& \frac{\alpha_1(k)(\omega_0(k)+\Omega_1(k))
    (\omega_0(k)^2-\Omega_2(k)^2)+\alpha_2(k)(\omega_0(k)+\Omega_2(k))
    (\omega_0(k)^2-\Omega_1(k)^2)}
        {2\omega_0(k)(\omega_0(k)^2-\omega_1(k)^2)}\\
    \tilde{\alpha}_2(k) &=& \frac{\alpha_1(k)(-\omega_0(k)+\Omega_1(k))
    (\omega_0(k)^2-\Omega_2(k)^2)+\alpha_2(k)(-\omega_0(k)+\Omega_2(k))
    (\omega_0(k)^2-\Omega_1(k)^2)}
        {-2\omega_0(k)(\omega_0(k)^2-\omega_1(k)^2)}\\
    \tilde{\alpha}_3(k) &=& \frac{\alpha_1(k)(\omega_1(k)+\Omega_1(k))
    (\omega_1(k)^2-\Omega_2(k)^2)+\alpha_2(k)(\omega_1(k)+\Omega_2(k))
    (\omega_1(k)^2-\Omega_1(k)^2)}
        {2\omega_1(k)(\omega_1(k)^2-\omega_0(k)^2)}\\
    \tilde{\alpha}_4(k) &=& \frac{\alpha_1(k)(-\omega_1(k)+\Omega_1(k))
    (\omega_1(k)^2-\Omega_2(k)^2)+\alpha_2(k)(-\omega_1(k)+\Omega_2(k))
    (\omega_1(k)^2-\Omega_1(k)^2)}
        {-2\omega_1(k)(\omega_1(k)^2-\omega_0(k)^2)}\\
    \tilde{\beta_1}(k) &=& \frac{\Delta_k(\omega_0(k)^2-
       (\alpha_1(k)\Omega_2(k)+\alpha_2(k)\Omega_1(k))^2)}
        {2\omega_0(k)(\omega_0^2-\omega_1^2)}\\
    \tilde{\beta_2}(k) &=& \frac{\Delta_k(\omega_0(k)^2-
       (\alpha_1(k)\Omega_2(k)+\alpha_2(k)\Omega_1(k))^2)}
        {2\omega_1(k)(\omega_1^2-\omega_0^2)} \;\;\; .
\end{eqnarray}

Knowing the spectral functions, we now insert them into the moment equations
(\ref{a0})--(\ref{a3}) and obtain the following for the
superconducting state:
\begin{eqnarray}
        \alpha_1(k)+\alpha_2(k) &=& 1 \\
        \alpha_1(k) \Omega_1(k) + \alpha_2(k) \Omega_2(k) &=& a_1 \\
        \alpha_1(k) \Omega_1^2(k) + \alpha_2(k) \Omega_2^2(k) &=&
                a_2 - \Delta_k^2 \\
        \alpha_1(k) \Omega_1^3(k) + \alpha_2(k) \Omega_2^3(k) &=&
                a_3 - \Delta_k^2 a_1 \;\;\; .
\end{eqnarray}
The first two off-diagonal moment equations, Eqs.\ (\ref{b0}) and (\ref{b1}),
are automatically satisfied by our Dyson's equation.

The resulting moment equations for the superconducting state can be
solved and uniquely determine the form of the spectral functions.
The second off-diagonal moment equation yields the gap equation
\begin{equation} \label{gabeq}
        \frac{\Delta_k}{U} = \sum_k \int_{-\infty}^\infty d\omega\,
        n_f(\omega) B(k,\omega) \;\;\; .
\end{equation}
 We immediately
see that in this approach the order parameter is a constant, $\Delta_k=\Delta$.
We now have to evaluate the order parameter, chemical potential,
band correction, $\alpha_1(k), \alpha_2(k), \Omega_1(k)$, and
$\Omega_2(k)$ self-consistently, using Eqs.\
(\ref{a0})--(\ref{a3}), (\ref{k-average}), (\ref{chempot}), and
(\ref{gabeq}).


\vfill\eject

\clearpage

\begin{references}

\bibitem{ding} Ding, H., Yokoja, T.,  Campuzano, J.C., Takahashi, T.,  
Randeria, M., Norman, M.R., Mochiku, T., Kadowaki, K., Giapintzakis, J.:
Nature {\bf 382}, 51 (1996).

\bibitem{loeser} Loeser, A.G., Shen, Z.\-X., Dessau, D.S., Marshall, D.S.,
Park, C.H., Fournier, P., Kapitulnik, A.: Science {\bf 273}, 325 (1996).

\bibitem{tmatrix}
        Micnas, R., Pedersen, M.H., Schafroth, S., Schneider, T., 
     Rodr\'{\i}guez-N\'u\~nez, J.J., Beck, H.: 
    Phys.\  Rev.\  B {\bf 52}, 16223 (1995)

\bibitem{BK}
    Kadanoff, L.P., Baym,  G.: {Quantum Statistical Mechanics}. Advanced
   Book Classics. New York: Addison-Wesley 1989;
    Baym, G., Kadanoff, L.P.: Phys.\  Rev.\ {\bf 124}, 287 (1961);
    Baym, G.: Phys.\  Rev.\  {\bf 127}, 1391 (1962)

\bibitem{Micnas_et_al}
        Micnas, R., Ranninger, J., and Robaszkiewicz, S.:
        Rev.\  Mod.\  Phys.\ {\bf 62}, 113 (1990)

\bibitem{Rodriguez}
         Rodr\'{\i}guez-N\'u\~nez, J.J., 
         Schafroth, S., Beck, H., Schneider, T.,
         Pedersen, M.H., Micnas, R.:
        Physica B {\bf 206--207}, 654 (1995)

\bibitem{Beckapp}
     Rodr\'{\i}guez-N\'u\~nez, J.J., Schafroth, S.,  Micnas, R., 
     Schneider, T.,  Beck, H., Pedersen, M.H.:
    J.\  Low Temp.\  Phys.\ {\bf 99}, 315 (1995)

\bibitem{schafroth} Schafroth, S., Rodr\'{\i}guez-N\'u\~nez, J.J.:
        Z.\ Phys.\ B, {\em in press}.

\bibitem{Nolting} Nolting, W.: Z.\ Physik {\bf 225}, 25 (1972);
        Nolting, W.: {Grundkurs: Theoretische
        Physik. 7 Viel-Teilchen-Theorie.}
        Ulmen: Zimmermann--Neufang 1992

\bibitem{moments} Schneider, T., Pedersen, M.H.,  
        Rodr\'{\i}guez-N\'u\~nez, J.J.:
        Z.\  Phys.\ B {\bf 100}, 263 (1996)

\bibitem{BW}
    Bickers, N.E., White, S.R.:
    Phys.\  Rev.\  B  {\bf 43}, 8044 (1991)

\bibitem{Hauss}
    Haussmann, R.; Z.\  Phys.\  B  {\bf 91}, 291 (1993)

\bibitem{Mattuck}
        Mattuck, R.D.: {A Guide to Feynman Diagrams
        in the Many-Body Problem}.  New York: Dover 1992;
        Eqs.\  (10.18) and  (15.58)

\bibitem{Fresard}
        Fr\'esard, R., Glaser, B., W\"olfle, P.:
        J.\  Phys.: Condens.\  Matter {\bf 4}, 8565 (1992)

\bibitem{M-R&F}
        Mart\'{\i}n--Rodero, A.,  Flores, F.:
        Phys.\  Rev.\  B {\bf  45},  13008 (1992)

\end{references}
\end{document}